
\def\epsfpreprint{Y}   % Turn on for preprint version with epsf figures

\def\draftversion{N}   % Turn on date and DRAFT across page
\def\preprint{Y}       % Turn on for preprint version

\input jnl \refstylenp \def\cit#1{[#1]}

\if \draftversion N \input reforder \input eqnorder \citeall\cit \fi

\if \preprint N \def\epsfpreprint{N} \fi
\if \epsfpreprint Y \input epsf \fi

\def\Fhi{\vec \Phi}  \def\half{{1 \over
2}} \def\fs{~~.}

%\ltapprox and \gtapprox produce > and < signs with twiddle underneath
\def\spose#1{\hbox to 0pt{#1\hss}}
\def\ltapprox{\mathrel{\spose{\lower 3pt\hbox{$\mathchar"218$}}
 \raise 2.0pt\hbox{$\mathchar"13C$}}}
\def\gtapprox{\mathrel{\spose{\lower 3pt\hbox{$\mathchar"218$}}
 \raise 2.0pt\hbox{$\mathchar"13E$}}}
\def\inapprox{\mathrel{\spose{\lower 3pt\hbox{$\mathchar"218$}}
 \raise 2.0pt\hbox{$\mathchar"232$}}}

\def\table#1#2{\halign{##\hfill\quad
&\vtop{\parindent=0pt \hsize=5.5in \strut## \strut}\cr {\bf Table
#1}&#2 \cr} }

\def\figure#1#2#3{\if \epsfpreprint Y \midinsert \epsfxsize=#3truein
\centerline{\epsffile{fig_#1_eps}} \halign{##\hfill\quad
&\vtop{\parindent=0pt \hsize=5.5in \strut## \strut}\cr {\bf Figure
#1}&#2 \cr} \endinsert \fi}

%                        FIGURE CAPTIONS
%%%%%%%%%%%%%%%%%%%%%%%%%%%%%%%%%%%%%%%%%%%%%%%%%%%%%%%%%%%%%%%%%%%%%%%

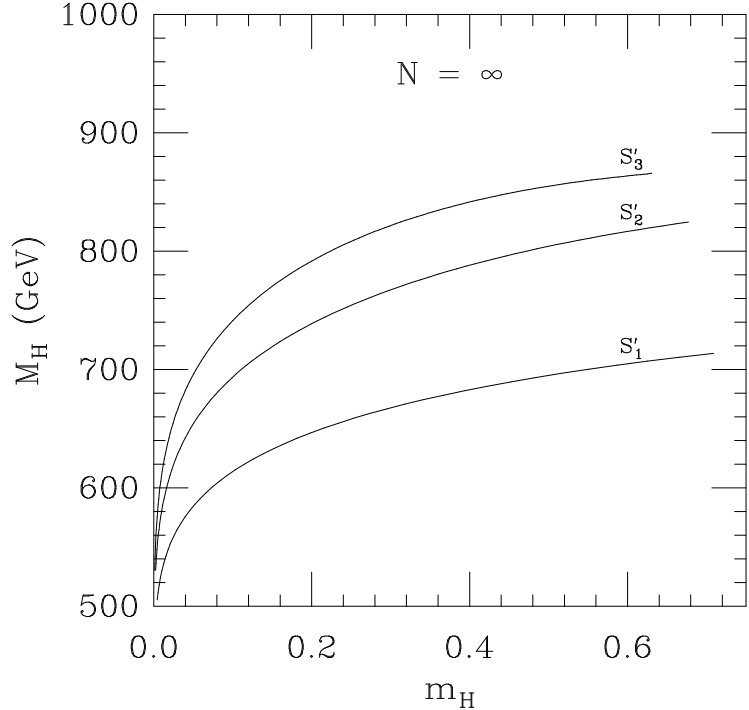
\def\captiontwoone{The Higgs mass $M_H = {{m_H}\over{f}} \times 246 ~
GeV$ in physical units vs. the Higgs mass $m_H$ in lattice units for
the three actions, eq. (2.1.2).}

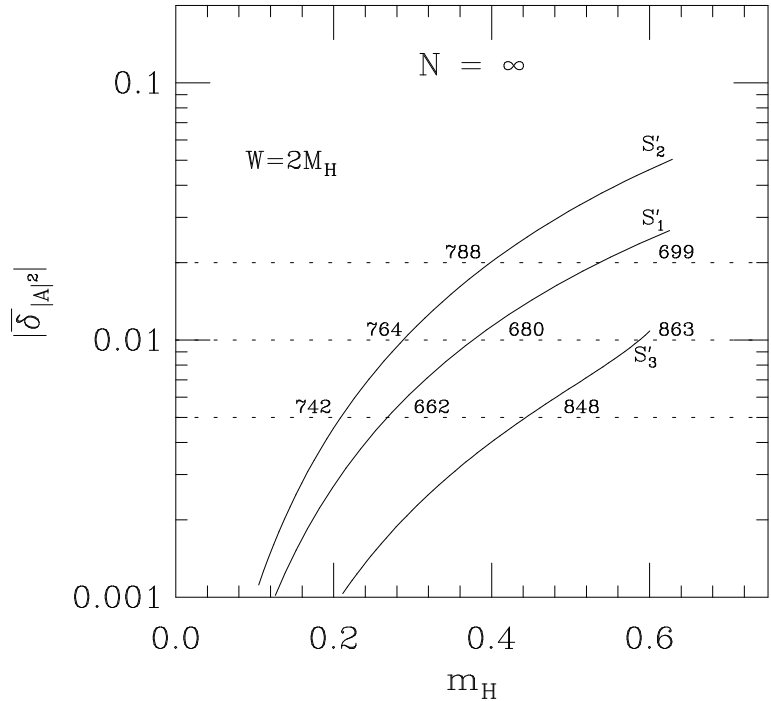
\def\captiontwotwo{Leading order cutoff effects in the invariant
$\pi-\pi$ scattering amplitude at $90^0$ at center of mass energy
$W=2M_H$ vs. the Higgs mass in lattice units for the three actions.
The values of $M_H$ in $GeV$ determined from $M_H = {{m_H}\over{f}}
\times 246 ~ GeV$ are put on the three horizontal lines at $\bar
\delta_{|A|^2} = 0.005,~0.01,~0.02$.}

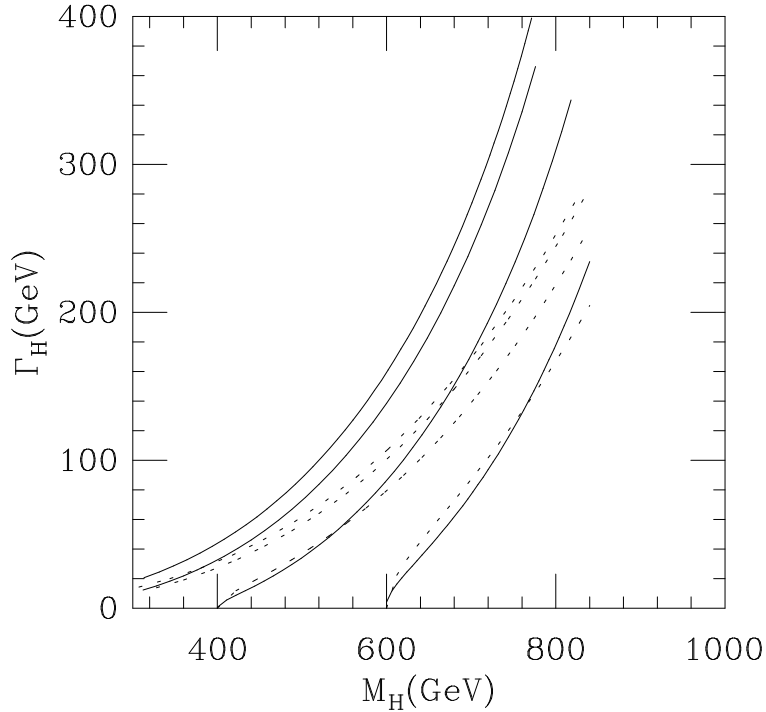
\def\captiontwothree{The universal part of the width vs. $M_H$.  The
solid line displays the large $N$ result scaled to $N=4$ and the
dotted line shows the leading order term in perturbation theory. From
left to right the lines correspond to pion masses $M_\pi =
0,~100,~200,$ and $300 ~ GeV$.}

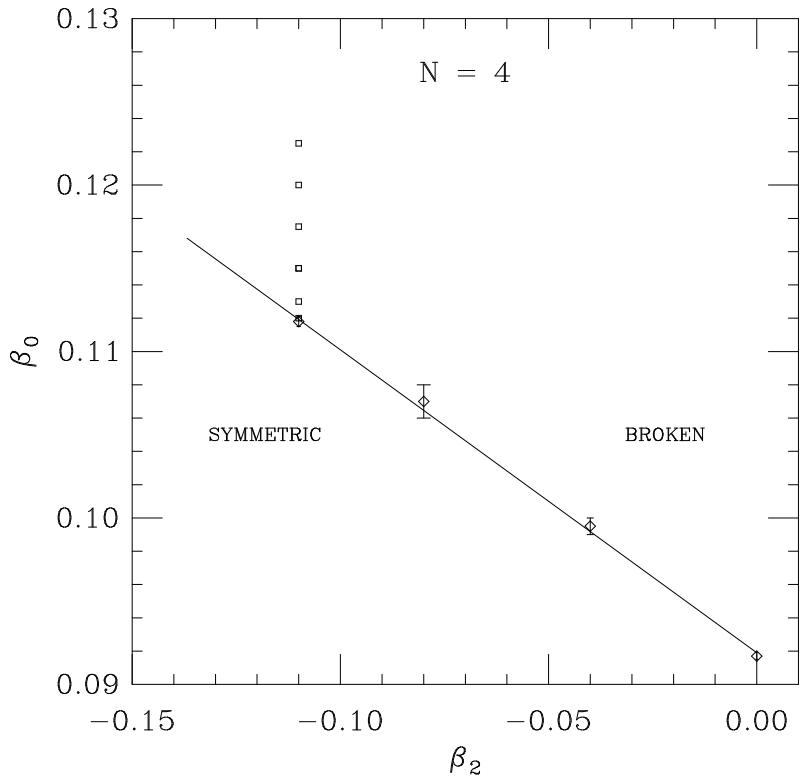
\def\captionthreeone{Phase diagram for action $S_2$. The solid line
is a least square straight line fit to the data that are denoted by
diamonds. The squares indicate the points where we made simulations to
determine $m_H$ and $f_\pi$. They all lie on the vertical line
$\beta_2 = -0.11$.}

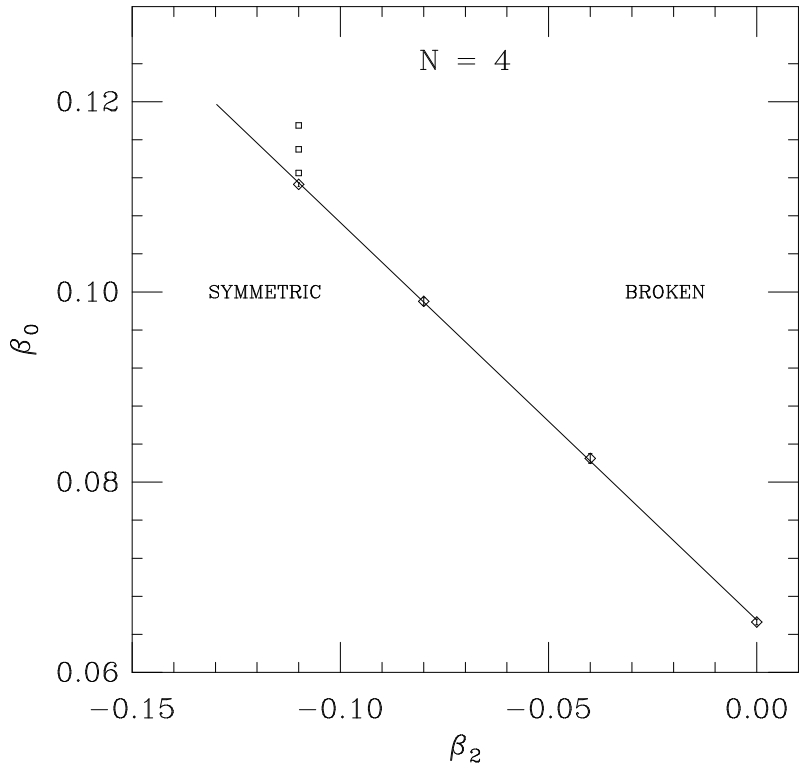
\def\captionthreetwo{Same as Figure 3.1 but for action $S_3$.}

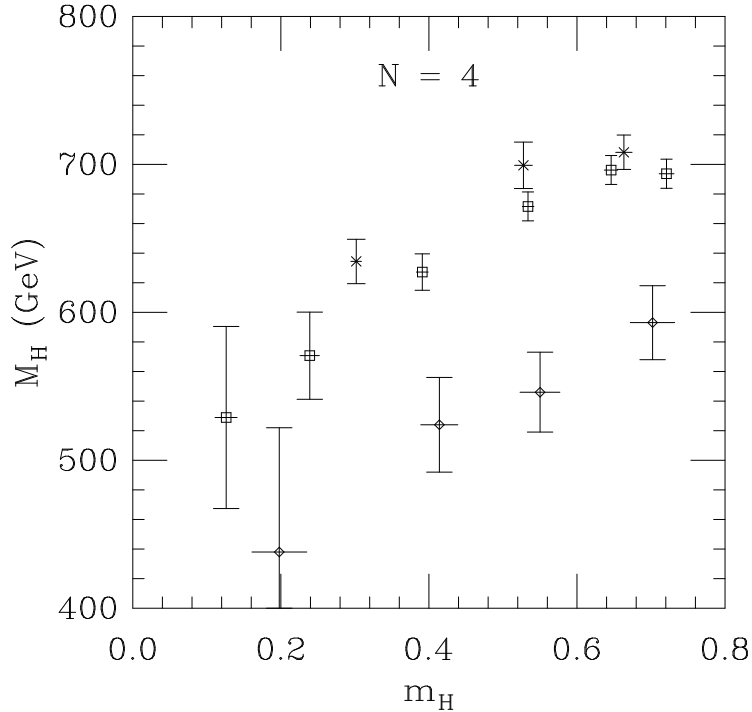
\def\captionthreethree{The Higgs mass $M_H = {{m_H}\over{f}} \times
246 ~ GeV$ in physical units vs. the Higgs mass $m_H$ in lattice units
from the numerical simulations. The diamonds correspond to action
$S_1$ \cit{BBHN2}, the squares to action $S_2$ and the crosses to
action $S_3$.}

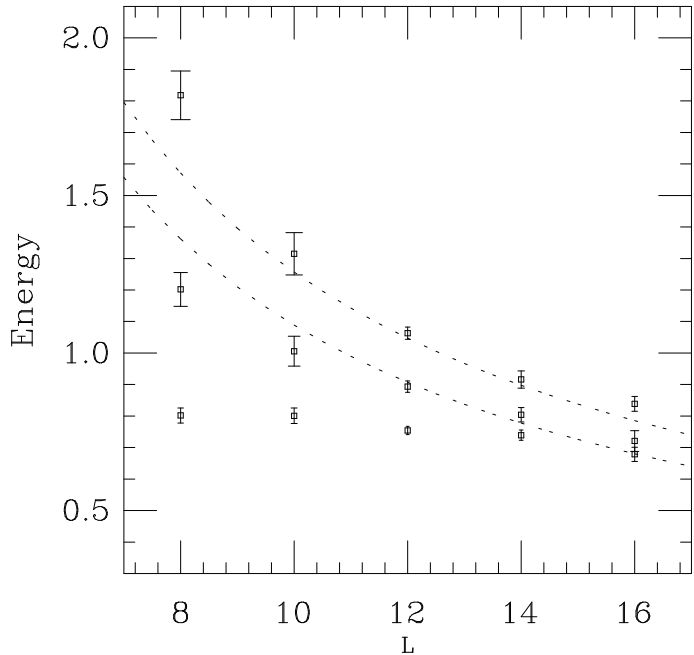
\def\captionthreefour{The low lying spectrum for action $S_1$ for
various lattice sizes as measured in a numerical simulation at
$\beta_0 = 0.10$. The dotted lines correspond to the two lowest
energies of two free pions.}

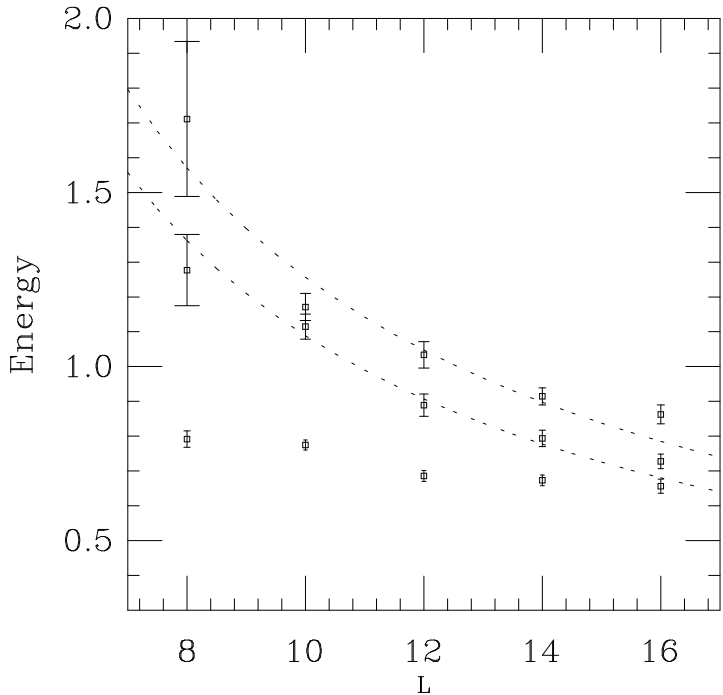
\def\captionthreefive{Same as Figure 3.4 but for action $S_2$ at
$\beta_0 = 0.12$, $\beta_2 = -0.11$.}

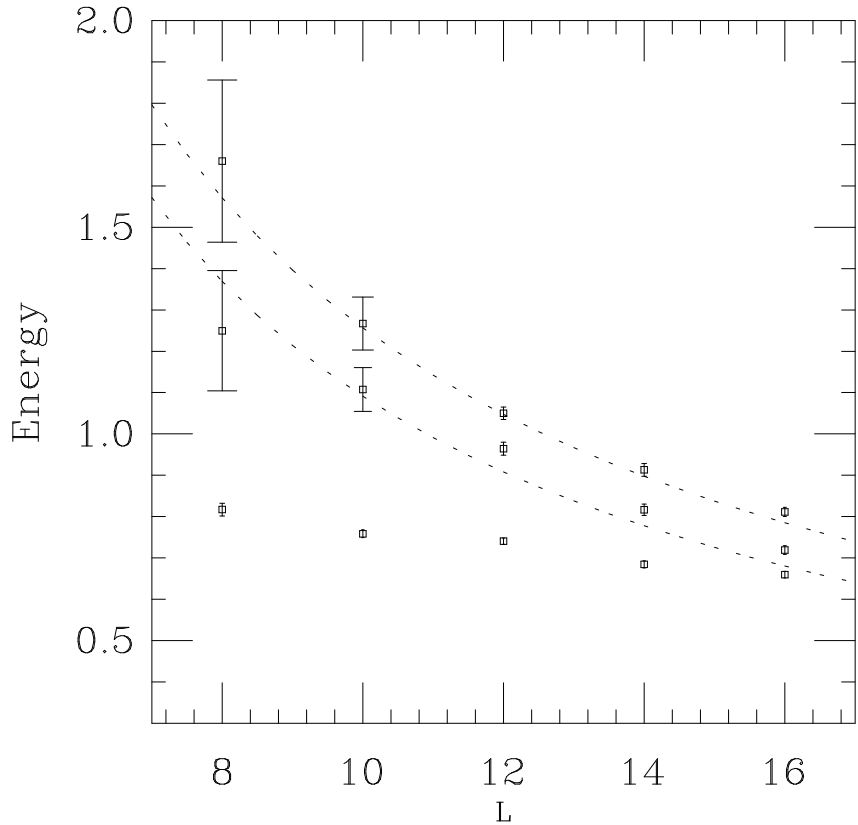
\def\captionthreesix{Same as Figure 3.4 but for action $S_3$
at $\beta_0 = 0.1175$, $\beta_2 = -0.11$.}

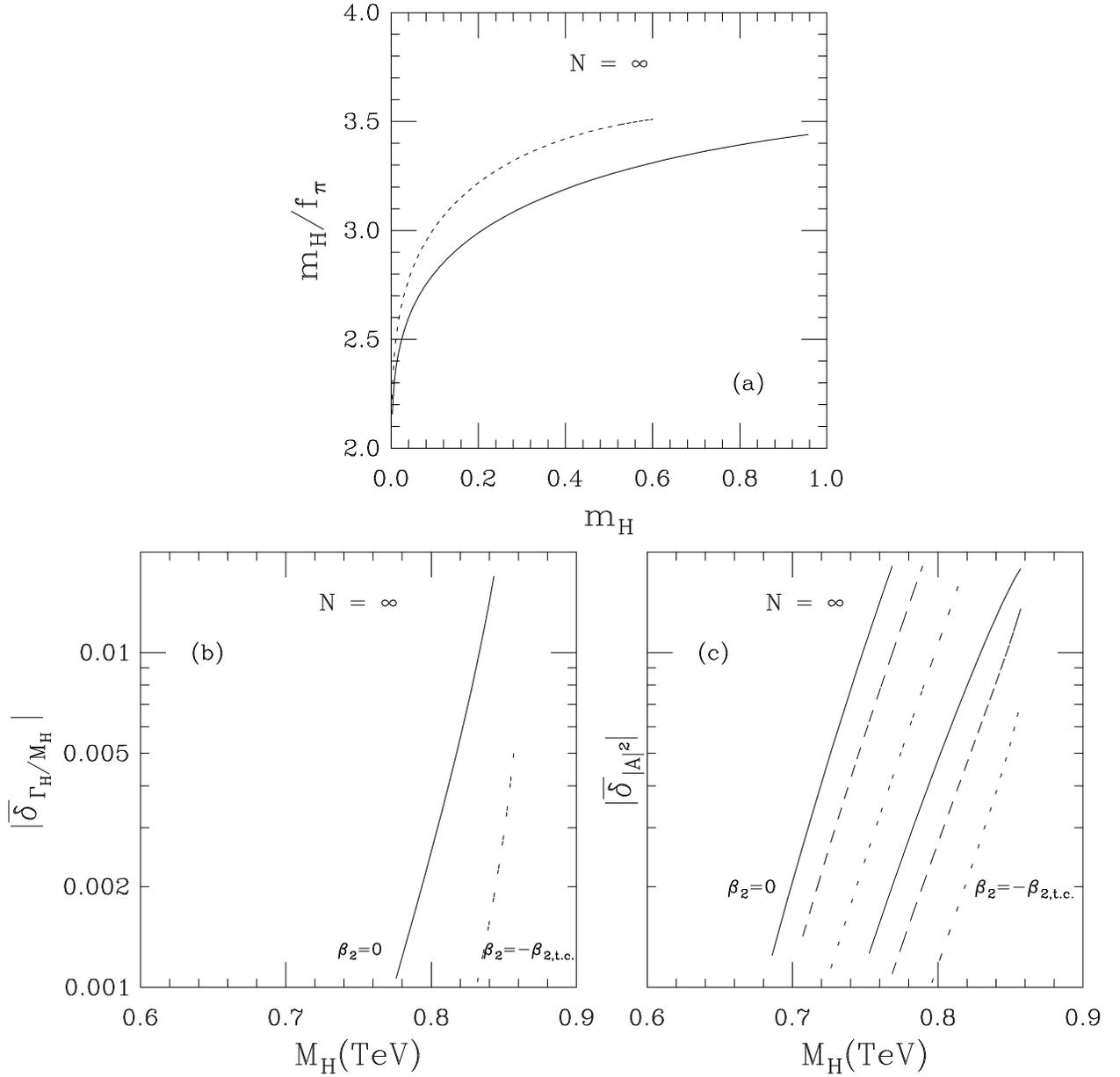
\def\captionAone{(a) $m_H/f_\pi$ vs. $m_H$: The solid line corresponds
to $\beta_2=0$ and the dotted line to the optimal value $\beta_2=
-\beta_{2,t.c.}$. (b) Leading order cutoff effect in the width to mass
ratio. (c) Leading order cutoff effects in the invariant $\pi-\pi$
scattering amplitude at $90^0$. Here the dotted line represents center
of mass energy $W=2M_H$, the dashed line $W=3M_H$ and the solid line
$W=4M_H$.}

%%%%%%%%%%%%%%%%%%%%%%%%%%%%%%%%%%%%%%%%%%%%%%%%%%%%%%%%%%%%%%%%%%%%%%%

%                       TABLE CAPTIONS
%%%%%%%%%%%%%%%%%%%%%%%%%%%%%%%%%%%%%%%%%%%%%%%%%%%%%%%%%%%%%%%%%%%%%%%
\def\tablecaptionone{Some points of the critical line for action $S_2$.}

\def\tablecaptiontwo{Some points of the critical line for action $S_3$.}

\def\tablecaptionthree{The Higgs mass $m_H$ in lattice units and the
Higgs mass $M_H = {{m_H}\over{f}} \times 246 ~ GeV$ in physical units
from the numerical simulations of action $S_1$. The errors quoted are
statistical errors only.}

\def\tablecaptionfour{Same as in Table 3 but for the action $S_2$ at
$\beta_2 = -0.11$.}

\def\tablecaptionfive{Same as in Table 3 but for the action $S_3$ at
$\beta_2 = -0.11$.}

\def\tablecaptionsix{The low lying spectrum for action $S_1$ for
various lattice sizes as measured in a numerical simulation at
$\beta_0 = 0.10$. The two lowest energy levels of a free two pion
state of zero total three--momentum are also given.}

\def\tablecaptionseven{Same as in Table 6 but for action $S_2$ at
$\beta_0 = 0.12$, $\beta_2 = -0.11$.}

\def\tablecaptioneight{Same as in Table 6 but for action $S_3$ at
$\beta_0 = 0.1175$, $\beta_2 = -0.11$.}
%%%%%%%%%%%%%%%%%%%%%%%%%%%%%%%%%%%%%%%%%%%%%%%%%%%%%%%%%%%%%%%%%%%%%%%

\if \draftversion Y


\fi

\def\today{\ifcase\month\or
	January\or February\or March\or April\or  May\or June\or
	July\or August\or September\or October\or November\or
	December\fi \space\number\day, \number\year}

\if \preprint Y \twelvepoint\oneandathirdspace \fi

\if \draftversion Y \rightline{\today{}} \fi

\if \preprint Y \rightline{FSU-SCRI-93-29} \fi
\if \preprint Y \rightline{CU-TP-590} \fi
\if \preprint Y \rightline{RU-93-06} \fi

\title Numerical analysis of the Higgs mass triviality bound

\vskip 1. cm

\author Urs M.~Heller${}^{a}$, Markus Klomfass${}^{b}$,
Herbert~Neuberger${}^{c}$,
Pavlos~Vranas${}^{a}$

\affil
\vskip 1.cm
\centerline{\it ${}^{a}$ Supercomputer Computations Research Institute}
\centerline{\it The Florida State University}
\centerline{\it Tallahassee, FL 32306}
\vskip .2cm
\centerline{\it ${}^{b}$ Department of Physics}
\centerline{\it Columbia University}
\centerline{\it New York, NY 10027}
\vskip .2cm
\centerline{\it ${}^{c}$ Department of Physics and Astronomy}
\centerline{\it Rutgers University}
\centerline{\it Piscataway, NJ 08855--0849}
\vskip 1.5 cm

\goodbreak

\abstract
Previous large $N$ calculations are combined with numerical work at
$N=4$ to show that the Minimal Standard Model will describe physics to
an accuracy of a few percent up to energies of the order 2 to 4 times
the Higgs mass, $M_H$, only if $M_H \le 710\pm60 ~ GeV$. This bound is
the result of a systematic search in the space of dimension six
operators and is expected to hold in the {\it continuum}. Given that
studying the scalar sector in isolation is already an approximation,
we believe that our result is sufficiently accurate and that further
refinements would be of progressively diminishing interest to particle
physics.

\endtitlepage

\if \preprint N \doublespace \fi

\head{1. Introduction and Conclusion.}
\taghead{1.}

Our goal is to obtain an estimate for the triviality bound on the
Higgs mass in the minimal standard model. Much work has preceded this
paper (for examples consult \cit{LQT,GENERAL,LW,GENERAL2,BBHN2,GOCK2}
and the review \cit{EINHBOOK}). We build on these results and
generalize them. Specifically, we deal more systematically with the
quantitative uncertainty resulting from the arbitrariness of the
coefficients of higher dimensional operators in the scalar sector
\cit{DALLAS}. By doing this we aim to obtain a number that is
meaningful beyond {\sl lattice} field theory and directly relevant to
particle physics.

The overall framework of the approach has been reviewed before (e.g.
\cit{DALLAS}) and will not be repeated here. Its main simplifying
feature is to treat the scalar sector of the minimal standard model in
isolation. When considering further efforts in this framework the
potential impact on particle physics should be evaluated against the
accuracy of neglecting other interactions, e.g.  with the top quark.
We believe that our result is reliable to a reasonable degree, given
that the whole framework is an approximation, and we feel that further
refinements would be of progressively diminishing interest to particle
physics\footnote{*} {An exception would be further investigations of
the Higgs width on the lattice.}.

Our result is that the minimal standard model will describe physics to
an accuracy of a few percent up to energies of the order 2 to 4 times
the Higgs mass, $M_H$, only if $M_H \le 710\pm 60 ~ GeV$. The two
major assumptions made are that ignoring all couplings but the scalar
self--coupling is a good approximation and that any higher energy
theory into which the standard model is embedded will not conspire to
eliminate all dimension six scalar field operators in the low energy
effective action. The number we obtain is not surprising because of
its closeness to tree level bounds \cit{LQT,LW}; what has been
achieved is to finally show that in any reasonable situation higher
orders in perturbation theory cannot change it substantially although
quite strong scalar self--interactions are possible.

In the sequel we shall rely quite heavily on \cit{LARGENLONG} but we
also try to make the paper accessible to readers who are not familiar
with \cit{LARGENLONG}. Our basic strategy was to first use ${1\over
N}$ expansion techniques in the generalization of the $O(4)$ symmetric
scalar sector to $O(N)$ to obtain an analytical non--perturbative
estimate for the bound and then follow up with Monte Carlo simulations
at the physical value $N=4$. In the next section we summarize needed
information at $N=\infty$ and add some new results in this limit. The
following section presents our numerical work with some emphasis on
the checks that were made to ascertain control over systematic errors.
The last section explains our main conclusion. Appendix A gives a few
more details on the new large $N$ results, and finally in appendix B
we collect the numbers obtained from our simulations in several tables
matching the graphs shown in the main text.

Our notational conventions are: The Higgs mass in physical units
($GeV$), defined as the real part of the resonance pole, is denoted by
$M_H$ and the width by $\Gamma_H$. The same quantities in lattice
units are denoted by lower case letters, $m_H$ and $\gamma_H$. The
matrix element of conventionally normalized currents of broken
symmetries between the vacuum and single Goldstone boson states
(referred to as pions, $\pi$) is denoted by $F$ in physical units and
by $f$ in lattice units. The scalar self--coupling is defined by:
$$
g={{3M_H^2}\over {F^2}} \fs \eqno(e1p1)
$$
In the $N=\infty$ section everything is written in terms of the above
$N=4$ notation.  We use $\Lambda$ to denote a generic cutoff.

Most of our work is on the $F_4$ lattice which can be thought of as
embedded in a hypercubic lattice from which odd sites (i.e. sites
whose integer coordinates add up to an odd sum) have been removed. The
lattice spacing of the hypercubic lattice is $a$.  It is set to unity
when $m_H$, $\gamma_H$ and $f$ are used. The $F_4$ lattice is always
fully symmetric having, when finite, $L$ sites in each principal axis
direction so that the total number of sites is $L^4$. Usually, $x,
x',x''$ denote sites, $<x,x'>,l,l'$ links, $\ll x,x'\gg$
next--nearest--neighboring pairs, $<l,l'>$ pairs of links, and the
field is constrained by ${\Fhi}^2 (x) =1$.

\head{2. Results at large $N$.}
\taghead{2.}

Here we summarize the large $N$ results of \cit{LARGENLONG} relevant
to our numerical work. They contain predictions that can be directly
compared to $N=4$ Monte Carlo data and evaluations of observable
cutoff effects on $\pi-\pi$ scattering. The dependence of the bound on
the magnitude of the observable cutoff effects is relatively
insensitive: a change by a factor of 3 induces a variation of $50 ~
GeV$ in the bound in the worst case. Thus, we do not worry about ${1
\over N}$ corrections to the cutoff effects. Cutoff effects could also
be calculated in perturbation theory but we have argued (see Appendix
B of \cit{LARGENLONG}) that the ${1 \over N}$ computation is probably
more reliable. Hence we use the ${1 \over N}$ results here.

\subhead{2.1. Relaxing the bound.}
\taghead{2.1.}

The na\"{\i}ve expectation that heavier Higgs masses are obtained when
the bare scalar self--coupling is increased is upheld by
nonperturbative calculations. The search for the bound can therefore
be restricted to nonlinear actions.

Among the nonlinear actions the bound is further increased by reducing
as much as possible the attraction between low momentum pions in the
$I=J=0$ channel. Given a bare action the approximate combination of
parameters achieving this is identified as follows. Expand in slowly
varying fields and use a field redefinition to bring the action,
including terms up to fourth order in the momenta, to the form
$$
S_c ~ = \int_x ~ \left [{1\over 2} \vec \phi ( -\partial^2 )
\vec \phi - {b_1 \over {2N}} (\partial_\mu \vec \phi
\cdot \partial_\mu \vec \phi )^2 - {b_2 \over {2N}} (\partial_\mu \vec
\phi \cdot \partial_\nu \vec \phi - {1\over 4} \delta_{\mu , \nu }
\partial_\sigma \vec \phi \cdot \partial_\sigma \vec \phi )^2
\right] \fs \eqno(e2p1p1)
$$
At $N=\infty$ $b_2$ has no effect and the bound depends monotonically
on $b_1$, increasing with decreasing $b_1$. Overall stability of the
homogeneous broken phase restricts the range of $b_1$. For example on
the $F_4$ lattice, at the optimal value of $b_1$ the bound is
increased by about $100 ~ GeV$ relative to the simplest non linear
action.

The rule in the above paragraph does not lead to an exactly universal
bound. Different bare actions that give the same effective parameter
$b_1$ can give somewhat different bounds because the dependence of
physical observables on the bare action is highly nonlinear. For
example, at the optimal $b_1$ value, Pauli--Villars regularizations
lead to bounds higher by about $100 ~ GeV$ than some lattice
regularizations. This difference between the lattice and
Pauli--Villars can be traced to the way the free massless inverse
Euclidean propagator departs from the $O(p^2 )$ behavior at low
momenta. For Pauli--Villars it bends upwards to enforce the needed
suppression of higher modes in the functional integral, while on the
lattice it typically bends downwards to reflect the eventual
compactification of momentum space.

Because we desire to preserve Lorentz invariance to order
$1/\Lambda^2$ we use the $F_4$ lattice and, on the basis of the above
observations, there are three stages of investigation. The first stage
is to investigate the na\"{\i}ve nearest--neighbor model. This should
be viewed as the generic lattice case where no special effort to
increase the bound is made. Since this case has been investigated
thoroughly in \cit{BBHN2} we can proceed to more complicated actions
with well tested methods of analysis. The next stage is to write down
the simplest action that has a tunable parameter $b_1$. The last stage
is to add a term to eliminate the ``wrong sign'' order $p^4$ term in
the free nearest--neighbor propagator, amounting to Symanzik
improvement of the large $N$ pion propagator. The three $F_4$ actions,
investigated at large $N$, are given by
$$
\eqalign{
S^\prime_1 =& ~ -2N\beta_0 \sum_{<x,x'>} \Fhi (x) \cdot \Fhi (x')\cr
S^\prime_2= & ~ S^\prime_1 ~-~{{N\beta_2 }\over 48} \sum_x
\left[\sum_{{l\cap x \ne
\emptyset}\atop {l=<x,x'>} } \Fhi (x) \cdot \Fhi (x') \right]^2 \cr
S^\prime_3 = & ~ - {N\beta_0} \sum_x \left[
2 \sum_{x' {\rm ~n.n.~to~}x } \Fhi (x) \cdot \Fhi (x') - {1 \over 2}
\sum_{x'' {\rm ~n.n.n.~to~}x } \Fhi (x) \cdot \Fhi (x'') \right] \cr
&~-~ {{N\beta_2 }\over 72} \sum_x \left[
2 \sum_{x' {\rm ~n.n.~to~}x } \Fhi (x) \cdot \Fhi (x') - {1 \over 2}
\sum_{x'' {\rm ~n.n.n.~to~}x } \Fhi (x) \cdot \Fhi (x'') \right]^2 \fs
\cr }\eqno(e2p1p2)
$$
These actions were chosen because they have relatively simple large
$N$ limits and permit a simultaneous study of both $F_4$ and
hypercubic lattices. In each case, at constant $\beta_2$, $\beta_0$ is
varied tracing out a line in parameter space approaching a critical
point from the broken phase. This line can also be parameterized by
$m_H$ or $g$. For actions $S^\prime_2$ and $S^\prime_3$, $\beta_2$ is
chosen so that on this line the bound on $M_H$ is expected to be
largest. A simulation produces a graph showing ${{m_H}\over{f}}$ as a
function of $m_H$ along this line. The y-axis is turned into an axis
for $M_H$ by $M_H = {{m_H}\over{f}} \times 246 ~ GeV$.  The large $N$
predictions for these graphs are shown in Figure 2.1. Since action
$S^\prime_3$ was not treated in \cit{LARGENLONG} we include a brief
account in appendix A.

\figure{2.1}{\captiontwoone}{4.8}

\subhead{2.2. Cutoff Effects.}
\taghead{2.2.}

At $N=\infty$ the cutoff effects to order $1/\Lambda^2$ on the Higgs
width and $\pi-\pi$ scattering are parameterizable by:
$$
S_{eff}=S_R (g) + c \exp [-96\pi^2 / g ] {\cal O}(g) \fs \eqno(e2p2p1)
$$
$S_R$ contains only universal information and so does ${\cal O}$. All
the non--universal information is in the $g$ independent parameter
$c$. The form of $S_{eff}$ reflects the factorization of cutoff
effects into a universal $g$ dependent function and a $g$ independent
non--universal amplitude. The bound is increased by first varying the
non--universal part so that $c$ decreases at constant $g$ and then
going along the selected line to higher $g$.

\figure{2.2}{\captiontwotwo}{4.8}

$\bar \delta_{|A|^2}$ denotes the fractional deviation of the square
of the $\pi^a -\pi^a $ ($a$ identifies one of the $N-1$ directions
transverse to the order parameter in internal space) scattering
amplitude at 90 degrees in the CM frame at energy $W$ from its large
$N$ universal value. A plot of $\bar \delta_{|A|^2}$ as a function of
$M_H (m_H )$ for the three actions is presented in Figure 2.2. If one
considers only the magnitude of cutoff effects as a function of $m_H
={{M_H} \over {\Lambda}}$, one might conclude that the bound obtained
with action $S^\prime_1$ would be larger than the bound obtained with
$S^\prime_2$. This conclusion proves to be wrong when the mass in
physical units is considered. The values of $M_H$ in $GeV$, determined
from $M_H = {{m_H}\over{f}} \times 246 ~ GeV$, are put on three
horizontal lines in Figure 2.2 at $\bar \delta_{|A|^2} =
0.005,~0.01,~0.02$. At large $N$ the bound increases when going from
$S^\prime_1$ to $S^\prime_2$ and then to $S^\prime_3$ by a little over
10\% at each step. For example, for $\bar\delta_{|A|^2} =.01$ we get
bounds on $M_H$ of 680, 764, 863 $GeV$ for $S^\prime_1$, $S^\prime_2$
and $S^\prime_3$ respectively.

\subhead{2.3. Width.}
\taghead{2.3.}

The width $\Gamma_H$ is important for phenomenology and for
lattice work (see sect. $3.4$). With massless pions perturbation
theory seriously underestimates $\Gamma_H$ when $M_H$ is large
\cit{LARGENLONG}.
\figure{2.3}{\captiontwothree}{4.8}
\noindent
It would therefore be desirable to determine the
width non--perturbatively. Up to date the only methods known for
measuring the width within a numerical simulation require a non--zero
pion mass \cit{LUSCHER}. To study the effects of a non--zero pion mass
we computed the width in this case at large $N$. The mass for the
pions was induced by an external magnetic field that breaks the
symmetry explicitly \cit{ZIMMERW}. Because the cutoff effects on the
width are very small we show in Figure 2.3 only the universal part and
compare it to the leading order perturbative values. Shown are the two
results for $M_\pi = 0,~100,~200,$ and $300 ~ GeV$.

We see that the deviations from perturbation theory decrease when the
pion mass increases. Therefore to detect nonperturbative effects on
the width at $N=4$ one would have to deal either with the massless
case directly or, at least, with quite light pions, say $M_\pi
\ltapprox {1\over 6} M_H$. This is one point we believe deserves
further study.

\head{3. The physical case $N=4$.}
\taghead{3.}

The primary aim of the numerical work is to produce at $N=4$ the
analogues of the graphs in Figure 2.1. The actions simulated are
slightly different than those investigated at large $N$. Firstly, the
factors $N$ accompanying the couplings in \(e2p1p2) are omitted.
Secondly, we take advantage of the fact that on the $F_4$ lattice,
unlike on the hypercubic lattice, the simplest action with a tunable
parameter $b_1$ can be constructed in a way that maintains the
nearest--neighbor character of the action. This is done by coupling
fields sited at the vertices of elementary bond--triangles. Finally we
add a term to eliminate the ``wrong sign'' order $p^4$ term in the
free nearest--neighbor propagator, leaving, unlike in $S^\prime_3$,
the term with the tunable $b_1$ coupling unchanged. This new term
couples next--nearest--neighbors and amounts to tree level Symanzik
improvement.\footnote{*}{On the hypercubic lattice tree level
improvement of the simplest action may also help reduce Lorentz
violation effects at order $1/\Lambda^2$ and was investigated in
\cit{GOCK2}.} The three $F_4$ actions simulated (with $S_i$ having the
same expansion in slowly varying fields as $S^\prime_i$ investigated
at large $N$) are
$$
\eqalign{
S_1 =& ~ -2\beta_0 \sum_{<x,x'>} \Fhi (x) \cdot \Fhi (x')\cr
S_2= & ~ S_1 ~-~{{\beta_2 }\over 8} \sum_x
\sum_{{{<ll'>}\atop {l,l' \cap  x \ne
\emptyset ,~ l\cap x' \ne\emptyset ,~ l'\cap x'' \ne\emptyset }}\atop
{x,x',x'' {}~{\rm all~ n.n.}} } \left [ \left( \Fhi (x) \cdot \Fhi (x')
\right) ~ \left(\Fhi (x) \cdot \Fhi (x'') \right) \right ] \cr
S_3 = & ~ - {2(2\beta_0 +\beta_2 )}\sum_{<x,x'>}  \Fhi (x) \cdot \Fhi (x')
{}~+ ~{(\beta_0 +\beta_2 )}\sum_{\ll x,x'\gg}  \Fhi (x) \cdot \Fhi (x') \cr
&~-~ {{\beta_2 }\over 8}\sum_x
\sum_{{{<ll'>}\atop {l,l' \cap  x \ne
\emptyset ,~ l\cap x' \ne\emptyset ,~ l'\cap x'' \ne\emptyset }}\atop
{x,x',x'' {}~{\rm all~ n.n.}} } \left [ \left( \Fhi (x) \cdot \Fhi (x')
\right) ~ \left(\Fhi (x) \cdot \Fhi (x'') \right) \right ] \fs \cr
}\eqno(e3p1)
$$
The data for action $S_1$ can be found in \cit{BBHN2}.  Preliminary
data for action $S_2$ were presented in \cit{DALLAS,AMSTERDAM} and
references therein. Our main task is to finalize the results for
action $S_2$ and carry out some new measurements for action $S_3$.

Each action is investigated in two main steps. First, the phase
diagram is established; next, a particular line is chosen in the
broken phase, which for actions $S_2$ and $S_3$, amounts to picking a
value for $\beta_2$. On this line we make several measurements at
different values of $\beta_0$ approaching the critical point
$\beta_{0c} (\beta_2 )$. We chose $\beta_2$ based on the large $N$
criteria and on the measured structure of the phase diagram. We may
therefore be missing the ``best'' action, but, from our experience at
large $N$, we expect at most a $20-30~GeV$ additional increase in
the bound. Because we object to excessive fine--tuning and since one
may view what we are doing already as some amount of fine tuning, the
$20-30~GeV$ might go in either direction and represents the
systematic uncertainty that we assign to the question of fine tuning.

\subhead{3.1. Methods in General.}
\taghead{3.1.}

We follow closely the approach of \cit{BBHN2}. We use a Metropolis
algorithm to map out the phase diagram and a single cluster spin
reflection algorithm, tested against the Metropolis algorithm, for the
actual measurements. Typically we use $10,000-100,000$ lattice passes,
depending on lattice size and couplings, and simulate systems of
increasing sizes with even $L$ (to avoid frustration effects),
$L=6,\cdots ,16$. The spin update speeds on a single processor CRAY
Y--MP are about $20{{\mu sec}\over {site}}$ for action $S_1$, $50{{\mu
sec}\over {site}}$ for $S_2$ and $55{{\mu sec}\over {site}}$ for
$S_3$.

The total computer time invested is $400~hrs$ for action $S_1$,
$500~hrs$ for $S_2$ and $400~hrs$ for $S_3$. The approximate constancy
of the total amount of time spent for each action, in spite of the
significant increase in complexity indicated by the rise of time per
spin update, reflects the diminishing need for checks as confidence in
the numerical methods builds up. The total amount of time spent,
approximately $1300~hrs$, is quite modest and shows that careful
preparation, continuous support by analytical work and an incremental
approach pay off.

Statistical errors are always treated by the Jackknife method and in
least $\chi^2$--fits correlations between measurements are taken into
account.

\subhead{3.2. Phase Diagrams.}
\taghead{3.2.}

For actions $S_2$ and $S_3$ we determine critical points $\beta_{0c}
(\beta_2 )$ at several fixed $\beta_2$'s using Binder cumulants of the
magnetization ${\cal M} = {{\sum_x \Fhi (x)}\over {L^4 }}$. The
critical points are obtained from the intersection points of the
Binder cumulant graphs for different volumes, produced by reweighting
and patching \cit{SWENDSEN} the measurements from several couplings
$\beta_0$. The results are shown in Figures 3.1 and 3.2 and the
actual numbers are listed in Tables $1$ and $2$. They compare well
with our large $N$ results which were helpful in deciding where to
scan in the first place. At large $N$ the critical lines are straight,
a feature that seems to persist at $N=4$.

\figure{3.1}{\captionthreeone}{4.8}
\figure{3.2}{\captionthreetwo}{4.8}

Based on our large $N$ work we know that we want extremal values of
$\beta_2$ \cit{LARGENLONG}. By this we mean the most negative value of
$\beta_2$ for which the leading term in the long wave--length
expansion, eq.~\(e2p1p1), still has the usual ferromagnetic sign on
the critical line ({\it i.e.} $\beta_{0c} + \beta_2 \gtapprox 0$). We
therefore select $\beta_2 =-0.11$ for actions $S_2$ and $S_3$. The
critical points at $\beta_2 =-0.11$ are $\beta_{0c} =0.1118(3)$ and
$\beta_{0c} =0.1113(2)$ respectively. The specific points we choose to
study in detail are shown in Figures 3.1 and 3.2.

\subhead{3.3. Coupling Constant.}
\taghead{3.3.}

To obtain the coupling constant $g$, eq.~\(e1p1), we need $f$ and
$m_H$. $f$ is obtained from measuring the magnetization, ${\cal M}$,
and the pion wave function renormalization constant, $Z_\pi$, defined
from the residue of the pole at zero momentum in the pion--pion
propagator.  The pion field, $\vec \pi$, is defined as the component
of $\Fhi$ transverse to ${\cal M}$, and for low momenta we have
$<|\vec\pi (p)|^2> = {{Z_\pi^2}\over{p^2}} + {\rm regular~terms}$.
The magnetization is obtained by extrapolating to $L=\infty$ the
quantities $<{\cal M}^2>_L$ with an $O(1/L^2 )$ correction, using the
methods in \cit{HELLN}. $f$ is then obtained from $f={\cal M}/Z_\pi$.
This method is safe and the reliability of the numbers one obtains for
$f$ is high for our purposes. The error never exceeds 1.5\% in the
region of higher Higgs masses which we are interested in. The estimate
of $f$ by analytical methods \cit{LW} and \cit{MK} has an error of
order 5\% in the same region; thus $f$ is better determined by Monte
Carlo.  This is due mainly to the good theoretical control one has
over finite volume effects \cit{BBHN2,GOCK2,HELLN}. Unfortunately, for
the determination of $m_H$ we are not so lucky.

To obtain $m_H$ we measure the correlations of zero total
three--momentum sigma states and low relative momentum two--pion
states at different time separations. The sigma field, $\sigma$, is
defined as the component of $\Fhi$ parallel to ${\cal M}$. $m_H$ is
the energy of one of the lightest states created by superpositions of
the $\sigma$ field and two--pion composite operators at zero total
three--momentum and is obtained from the eigenvalues of the measured
time correlation matrix. The evaluation of $m_H$ is not as clean as
that of $f$ and the next subsection discusses the determination of the
numbers in greater detail.

The main result of this paper is in Figure 3.3 which shows $M_H
=246\sqrt{g/3} ~GeV$ as a function of $m_H$ for all three actions. The
actual numbers are given in Tables $3$, $4$ and $5$. One clearly sees
the progressive increase of the bound. A glance at Figure 2.2 shows
that in all cases the cutoff effects on the pion--pion scattering are
below a few percent even at the maximal $M_H$ of each curve. Thus we
can take the largest of these maxima as our bound. The ordering of the
points and their relative positions are in agreement with Figure 2.1,
while the differences in overall scale, reflecting the difference
between $N=\infty$ and $N=4$, come out as expected \cit{LARGENLONG}.

\figure{3.3}{\captionthreethree}{4.8}

\subhead{3.4. More about mass determination.}
\taghead{3.4.}

The main source of systematic errors is in the evaluation of $m_H$. In
an infinite volume the Higgs particle would decay into two pions and
extracting $m_H$ from the fall--off of the $\sigma$--$\sigma$
correlation function would yield nonsense. In a finite volume the
decay is prohibited or severely restricted by the rather large minimal
amount of energy even the softest pions have due to momentum
quantization. This makes the measurement possible but leaves one with
the difficult task of estimating the accuracy of the so determined
real part of the resonance pole.

In \cit{BBHN2} we dealt with action $S_1$. As a check of the results
obtained in the broken phase we also measured the coupling $g$ in the
symmetric phase (as defined from the zero four--momentum four--point
correlation) and used perturbation theory to predict $g$ in the broken
phase. This method is free of finite width contamination and gave
results consistent with the broken phase analysis.  However, the
determination of the coupling in the symmetric phase had a large
statistical error. This problem has been eliminated by one of us
\cit{MK} who carried out a complete analysis modeled on the work of
\cit{LW}. The numbers of \cit{BBHN2} survive this test reasonably.  Of
course, the analytical method suffers from systematic errors of a
different type. It is difficult to know how much of the discrepancies
at larger mass values in the broken phase between the analytical and
the Monte Carlo results are due to either approach. Still, the
comparison gives a worst--case scenario for the systematics in both
cases, since the methods are so different that systematics are
unlikely to conspire to work in the same direction.  Indeed, while the
discrepancies on the $F_4$ lattice and on the hypercubic lattice
\cit{GOCK2} are roughly of similar magnitude, the sign is opposite. We
cannot be very precise, but the general conclusion we draw is that for
${{m_H}\over f}$ as a function of $m_H$ the overall systematic error
in the Monte Carlo results cannot exceed a few percent for the highest
Higgs masses measured.

For $S_2$ and $S_3$ an analysis similar to \cit{MK} would be very
demanding and has not been carried out.  In \cit{BBHN2} and for the
lighter Higgs masses of actions $S_2$ and $S_3$ the Higgs mass was
extracted using only the $\sigma$ field correlation function. Since
now, for the larger masses, the width is suspected to be larger some
additional checks are needed.  The most direct test is to write down
reasonable operators that would create predominantly two pion states
and check for mixing effects. We did this for action $S_1$ at
$\beta_0=0.10$, the coupling from which the bound quoted in
\cit{BBHN2} was extracted, and also for actions $S_2$ and $S_3$ at the
couplings where we extract the bound. The large $N$ results of section
2.3 show that more ambitious attempts to actually measure the width on
the lattice by making the pions massive \cit{ZIMMERW} may not be
directly relevant to the massless case.

The mass is now estimated by measuring the correlation matrix
${\cal C}(t)$
$$
{\cal C}_{ij}(t) = < {\cal O}_i(t) {\cal O}_j(0) > - <{\cal
O}_i(t)> <{\cal O}_j(0)>
$$
between several operators, ${\cal O}_i$, described in the previous
subsection, evaluated on two time slices separated by $t$ lattice
spacings. The eigenvalues are determined by diagonalizing ${\cal
C}(t_0)^{-1/2} ~{\cal C}(t) ~{\cal C}(t_0)^{-1/2}$ as in \cit{LWFF}
with $t_0=1$ except for the largest lattice with action $S_1$ where
$t_0=0$ was used. After some trials it was decided to diagonalize a
$3\times 3$ matrix. For each lattice size we evaluate the lowest
eigenvalue with this method. We also compute a ``trial mass" from the
$\sigma$ field correlation function alone. When the two numbers are
different within errors we expect to have some level repulsion between
the two lowest eigenvalues. We then identify the eigenvalue closer to
the free two pion energy as the energy of the two pion state and the
other as the resonance energy. To approximately correct for the
repulsion we adjust the resonance energy by the amount that the two
pion energy differs from its free value.  This technique should
eliminate leading $1/L$ corrections when they are significant
numerically. The so obtained Higgs masses are extrapolated to
$L=\infty$ by assuming an ${1\over{L^2}}$ behavior.  The quality of
these fits is acceptable but the motivation for the method of analysis
is somewhat empirical and therefore we allow for an order 3\%
systematic error in the mass determination.

\figure{3.4}{\captionthreefour}{4.6}

The low lying spectrum for selected high mass values for each action
are shown in Figures 3.4, 3.5 and 3.6 with the actual numbers
given in Tables $6$, $7$ and $8$. The new method of analysis confirms
that the older results for action $S_1$ are acceptable, which in turn
increases our confidence when the method is applied to actions $S_2$
and $S_3$. We see no indications for strong mixing because the largest
level repulsions observed are small compared to the eigenvalues
themselves. The two--pion states have a spectrum well described by the
lattice free particle dispersion and show only small amounts of
sensitivity to the resonance.  Taking into account statistical errors
we settle on an error estimate on the mass determinations of 5\%. This
results in an error of about 6\% on the determination of the physical
$M_H$.

\figure{3.5}{\captionthreefive}{4.8}
\figure{3.6}{\captionthreesix}{5.7}

\head{4. Main Result.}

A realistic and not overly conservative value for the Higgs mass
triviality bound is $710~GeV$. At $710~GeV$ the Higgs particle is
expected to have a width of about $210~GeV$ and is therefore quite
strongly interacting. We estimate the overall accuracy of our bound as
$\pm 60~GeV$. This includes the statistical error and systematic
uncertainty of the measurements of $M_H$ (see section 3.4) as well as
the systematic uncertainty assigned to fine tuning. The latter should
also allow for effects of the hereto neglected $b_2$ coupling in
eq.~\(e2p1p1). The meaning of the error is that it would be quite
surprising if evidence were produced for a reasonable generic model of
the scalar sector with observable cutoff effects bound by a few
percent, but a Higgs mass larger than $770~GeV$. It would be even more
surprising if a future analysis ended up concluding that the bound is
some number less than $650~GeV$.

The first estimates for the nearest--neighbor hypercubic actions gave
a bound of $640~GeV$ \cit{LW,GENERAL2} which is about 10\% below the
new number whereas the old $F_4$ bound was $590 ~GeV$ \cit{BBHN2}.
Thus, the older results have proven to be quite robust and this is an
indication that more search in the space of actions is unlikely to
yield surprises.

\head{Acknowledgments.}
The simulations for obtaining the phase diagram and measurements on
some of the smaller lattices were done on the cluster of IBM
workstations at SCRI. The other measurements were done on the CRAY
Y-MP at FSU. This research was supported in part by the DOE under
grant \# DE-FG05-85ER\discretionary{}{}{}250000 (UMH, MK and PV),
grant \# DE-FG05-92ER40742 (UMH and PV), and under grant \#
DE-FG05-90ER40559 (HN).

\vskip 1. cm

\head{Appendix A.}
\taghead{A.}

For the interested reader we sketch in this appendix the large $N$
analysis for action $S^\prime_3$ that was not treated in
\cit{LARGENLONG}. We can write the action $S^\prime_3$ as in eq.~(5.1)
of \cit{LARGENLONG}
$$
S^\prime_3 = \eta N (\beta_0 + \beta_2) \half \int_{x,y}
\Fhi (x) g_{x,y} \Fhi (y) - N \beta_2 {\eta^2 \over {8\epsilon} }
\int_x \left[ \int_y \Fhi (x) g_{x,y} \Fhi (y) \right]^2
\eqno(Ap1)
$$
with
$$
\eqalign{
\eta = & ~ 6 ~,~~~ \epsilon = 18 ~,~~~ \int_x = 2 \sum_x {}~,~~~
\int_p = \int_{B^*} {d^4 p \over {(2 \pi)^4}} \cr
g_{x,y} = & ~ 6 \tilde \delta_{x,y} - {1 \over 3}
\sum_{x' {\rm ~n.n.~to~}x } \tilde \delta_{y,x'} + {1 \over 12}
\sum_{x'' {\rm ~n.n.n.~to~}x } \tilde \delta_{y,x''} \cr
g(p) = & ~ {1 \over 3} \sum_{\mu \ne \nu} \left[ 2 - \cos(p_\mu
+ p_\nu) - \cos(p_\mu - p_\nu) \right] \cr
& ~ - {1 \over 12} \left\{ \sum_\mu \left[ 2 - 2\cos(2p_\mu) \right] +
\sum_{ \{ \epsilon_\mu = \pm 1 \} } \left[ 1 - \cos\left( \sum_\mu
\epsilon_\mu p_\mu \right) \right] \right\} \cr
= & ~ p^2 + O(p^6) \fs \cr
} \eqno(Ap2)
$$
Here we used $\tilde \delta_{x,y} = 1/2 \delta_{x,y}$ for an $F_4$
lattice and $B^*$ is its Brillouin zone.

With the above notations the computations of the phase diagram, Higgs
mass and cutoff effects are just as in sections 5 and 6 of
\cit{LARGENLONG}. We only need a few numerical constants. They are
defined in \cit{LARGENLONG} and for the present case take the values
$$
\eqalign{
r_0 = & ~ 0.09932603 \cr
c_1 = & ~ 0.0281844 ~,~~ c_2 = 2.2639 \cdot 10^{-4} \cr
\gamma = & ~ 0.01089861 \cr
\zeta = & 0 \fs \cr
} \eqno(Ap3)
$$
The resulting large $N$ plots for the Higgs mass and cutoff effects,
analogous to those in \cit{LARGENLONG} for the actions considered
there, are shown in Figure~A.1.

\figure{A.1}{\captionAone}{6.7}

\head{Appendix B.}

In this appendix we collect the numbers obtained from our simulations
in several tables matching the graphs shown in the main text.

%table 1
$$
\vbox{\tabskip=0pt \offinterlineskip
\def\tablerule{\noalign{\hrule}}
\halign to250pt{\strut#&\vrule height 12pt #\tabskip=1em plus2em &
\hfil#&\vrule# &
#\hfil&\vrule#\tabskip=0pt\cr\tablerule
&&\omit\hidewidth $\beta_2$ \hidewidth&&
\omit\hidewidth $\beta_{0c}$ \hidewidth&\cr\tablerule
&&$  0.000 $ && $ 0.0917(~2) $ &\cr
&&$ -0.040 $ && $ 0.0995(~5) $ &\cr
&&$ -0.080 $ && $ 0.1070(10)$ &\cr
&&$ -0.110 $ && $ 0.1118(~3) $ &\cr\tablerule }}
$$
\table{1}{\tablecaptionone}

%table 2
$$
\vbox{\tabskip=0pt \offinterlineskip
\def\tablerule{\noalign{\hrule}}
\halign to250pt{\strut#&\vrule height 12pt #\tabskip=1em plus2em &
 \hfil# & \vrule# &
 #\hfil & \vrule#\tabskip=0pt\cr\tablerule
&&\omit\hidewidth $\beta_2$ \hidewidth&&
  \omit\hidewidth $\beta_{0c}$ \hidewidth&\cr\tablerule
&&$  0.000 $ && $ 0.0653(3) $ &\cr
&&$ -0.040 $ && $ 0.0825(5) $ &\cr
&&$ -0.080 $ && $ 0.0990(4) $ &\cr
&&$ -0.110 $ && $ 0.1113(2) $ &\cr\tablerule }}
$$
\table{2}{\tablecaptiontwo}

%table 3
$$
\vbox{\tabskip=0pt \offinterlineskip
\def\tablerule{\noalign{\hrule}}
\halign to250pt{\strut#&\vrule height 12pt #\tabskip=1em plus2em &
\hfil#&\vrule#&\hfil#&\vrule#&
#\hfil & \vrule#\tabskip=0pt\cr\tablerule
&&\omit\hidewidth $\beta_0$ \hidewidth&&
\omit\hidewidth $m_H$ \hidewidth&&
\omit\hidewidth $M_H ~(GeV)$ \hidewidth&
\cr\tablerule
&&$ 0.0925 $ && $ 0.198(37) $ && $ 438(84) $ &\cr
&&$ 0.0950 $ && $ 0.414(25) $ && $ 524(32) $ &\cr
&&$ 0.0975 $ && $ 0.550(27) $ && $ 546(27) $ &\cr
&&$ 0.1000 $ && $ 0.702(30) $ && $ 593(25) $ &\cr
&&$ 0.1050 $ && $ 0.830(41) $ && $ 576(30) $ &\cr
&&$ 0.1100 $ && $ 0.946(50) $ && $ 568(30) $ &\cr
\tablerule }}
$$
\table{3}{\tablecaptionthree}

%table 4
$$
\vbox{\tabskip=0pt \offinterlineskip
\def\tablerule{\noalign{\hrule}}
\halign to250pt{\strut#&\vrule height 12pt #\tabskip=1em plus2em &
\hfil#&\vrule#&\hfil#&\vrule#&
#\hfil & \vrule#\tabskip=0pt\cr\tablerule
&&\omit\hidewidth $\beta_0$ \hidewidth&&
\omit\hidewidth $m_H$ \hidewidth&&
\omit\hidewidth $M_H ~(GeV)$ \hidewidth&
\cr\tablerule
&&$ 0.1120 $ && $ 0.126(15) $ && $ 529(62) $ &\cr
&&$ 0.1130 $ && $ 0.239(13) $ && $ 571(30) $ &\cr
&&$ 0.1150 $ && $ 0.391(~8) $ && $ 627(12) $ &\cr
&&$ 0.1175 $ && $ 0.534(~8) $ && $ 672(10) $ &\cr
&&$ 0.1200 $ && $ 0.646(~9) $ && $ 696(10) $ &\cr
&&$ 0.1225 $ && $ 0.721(10) $ && $ 694(10) $ &\cr
\tablerule }}
$$
\table{4}{\tablecaptionfour}

%table 5
$$
\vbox{\tabskip=0pt \offinterlineskip
\def\tablerule{\noalign{\hrule}}
\halign to250pt{\strut#&\vrule height 12pt #\tabskip=1em plus2em &
\hfil#&\vrule#&\hfil#&\vrule#&
#\hfil & \vrule#\tabskip=0pt\cr\tablerule
&&\omit\hidewidth $\beta_0$ \hidewidth&&
\omit\hidewidth $m_H$ \hidewidth&&
\omit\hidewidth $M_H ~(GeV)$ \hidewidth&
\cr\tablerule
&&$ 0.1125 $ && $ 0.302(~7) $ && $ 634(15) $ &\cr
&&$ 0.1150 $ && $ 0.528(12) $ && $ 699(16) $ &\cr
&&$ 0.1175 $ && $ 0.663(11) $ && $ 708(12) $ &\cr
\tablerule }}
$$
\table{5}{\tablecaptionfive}

%table 6
$$
\vbox{\tabskip=0pt \offinterlineskip
\def\tablerule{\noalign{\hrule}}
\halign to\hsize {\strut#&\vrule height 12pt#\tabskip=0em plus4em &
\hfil#&\vrule#\hskip0.1cm \vrule&
\hfil#&\vrule#&\hfil#&\vrule#&\hfil#&
\vrule#\hskip0.1cm \vrule&
\hfil#&\vrule#&\hfil#&\vrule#\tabskip=0pt\cr
\tablerule
&&L&&\multispan5\hidewidth Three lowest energy levels\hidewidth &&
\multispan3\hidewidth Two lowest free $\pi-\pi$ \hidewidth&
\cr\tablerule
&& $~8$ && $0.802(24)$ && $1.202(54)$ && $1.818(77)$ &&
 $1.362$ && $1.571$&\cr
&& $10$ && $0.801(25)$ && $1.005(47)$ && $1.315(67)$ &&
 $1.089$ && $1.257$&\cr
&& $12$ && $0.754(13)$ && $0.894(18)$ && $1.063(19)$ &&
 $0.907$ && $1.047$&\cr
&& $14$ && $0.739(16)$ && $0.804(23)$ && $0.916(27)$ &&
 $0.777$ && $0.898$&\cr
&& $16$ && $0.721(33)$ && $0.679(23)$ && $0.839(23)$ &&
 $0.680$ && $0.785$&\cr
\tablerule }}
$$
\table{6}{\tablecaptionsix}

%table 7
$$
\vbox{\tabskip=0pt \offinterlineskip
\def\tablerule{\noalign{\hrule}}
\halign to\hsize {\strut#&\vrule height 12pt#\tabskip=0em plus4em &
\hfil#&\vrule#\hskip0.1cm \vrule&
\hfil#&\vrule#&\hfil#&\vrule#&\hfil#&
\vrule#\hskip0.1cm \vrule&
\hfil#&\vrule#&\hfil#&\vrule#\tabskip=0pt\cr
\tablerule
&&L&&\multispan5\hidewidth Three lowest energy levels\hidewidth &&
\multispan3\hidewidth Two lowest free $\pi-\pi$ \hidewidth&
\cr\tablerule
&& $~8$ && $0.792(24)$ && $1.277(102)$ && $1.711(223)$ &&
 $1.362$ && $1.571$&\cr
&& $10$ && $0.775(14)$ && $1.115(~36)$ && $1.171(~39)$ &&
 $1.089$ && $1.257$&\cr
&& $12$ && $0.686(16)$ && $0.889(~32)$ && $1.034(~38)$ &&
 $0.907$ && $1.047$&\cr
&& $14$ && $0.673(15)$ && $0.794(~24)$ && $0.914(~25)$ &&
 $0.777$ && $0.898$&\cr
&& $16$ && $0.656(20)$ && $0.728(~21)$ && $0.863(~27)$ &&
 $0.680$ && $0.785$&\cr
\tablerule }}
$$
\table{7}{\tablecaptionseven}

%table 8
$$
\vbox{\tabskip=0pt \offinterlineskip
\def\tablerule{\noalign{\hrule}}
\halign to\hsize {\strut#&\vrule height 12pt#\tabskip=0em plus4em &
\hfil#&\vrule#\hskip0.1cm \vrule&
\hfil#&\vrule#&\hfil#&\vrule#&\hfil#&
\vrule#\hskip0.1cm \vrule&
\hfil#&\vrule#&\hfil#&\vrule#\tabskip=0pt\cr
\tablerule
&&L&&\multispan5\hidewidth Three lowest energy levels\hidewidth &&
\multispan3\hidewidth Two lowest free $\pi-\pi$ \hidewidth&
\cr\tablerule
&& $~8$ && $0.817(15)$ && $1.250(146)$ && $1.660(196)$ &&
 $1.369$ && $1.571$&\cr
&& $10$ && $0.759(~9)$ && $1.108(~53)$ && $1.267(~64)$ &&
 $1.091$ && $1.257$&\cr
&& $12$ && $0.740(~8)$ && $0.964(~16)$ && $1.050(~15)$ &&
 $0.908$ && $1.047$&\cr
&& $14$ && $0.684(~8)$ && $0.817(~14)$ && $0.913(~15)$ &&
 $0.778$ && $0.898$&\cr
&& $16$ && $0.659(~8)$ && $0.719(~10)$ && $0.811(~11)$ &&
 $0.680$ && $0.785$&\cr
\tablerule }}
$$
\table{8}{\tablecaptioneight}

\references

\refis{LQT} B. W. Lee, C. Quigg, H. B. Thacker, \pr D16, 1977, 1519.

\refis{GENERAL} N. Cabbibo, L. Maiani, G. Parisi, R. Petronzio, \np
B158, 1979, 295; B. Freedman, P. Smolensky, D. Weingarten, \pl 113B,
1982, 209; R. Dashen and H. Neuberger, \prl 50, 1983, 1897; H.
Neuberger, \pl B199, 1987, 536.

\refis{LW} M. L\"{u}scher and P. Weisz, \np B318, 1989, 705.

\refis{GENERAL2} J. Kuti, L. Lin and Y. Shen, \prl 61, 1988, 678; A.
Hasenfratz, K. Jansen, J. Jersak, C. B. Lang, T. Neuhaus, H. Yoneyama
\np B317, 1989, 81; G. Bhanot, K. Bitar, \prl 61, 1988, 798; U. M.
Heller, H. Neuberger and P. Vranas \pl B283, 1992, 335; U. M. Heller,
M. Klomfass, H. Neuberger and P. Vranas, \np B (Proc. Suppl) 26, 1992,
522.

\refis{BBHN2} G. Bhanot, K. Bitar, U. M. Heller and H. Neuberger, \np
B353, 1991, 551 ({\sl Erratum} \np B375, 1992, 503).

\refis{GOCK2} M. G{\"o}ckeler, H. A. Kastrup, T. Neuhaus and F.
Zimmermann, preprint HLRZ 92--35/PITHA 92-21.

\refis{DALLAS} H. Neuberger, U. M. Heller, M. Klomfass, P. Vranas,
Talk delivered at the XXVI International Conference on High Energy
Physics, August 6-12, 1992, Dallas, TX, USA; FSU--SCRI--92C--114, to
appear in the proceedings.

\refis{EINHBOOK} ``The Standard Model Higgs Boson'', {\bf Vol. 8},
{\sl Current Physics Sources and Comments}, edited by M. B. Einhorn,
North Holland, 1991.

\refis{LARGENLONG} U. M. Heller, H. Neuberger, P. Vranas,
FSU--SCRI--92--99/RU--92--18, to appear in {\sl Nucl. Phys.} {\bf B}.

\refis{AMSTERDAM} U. M. Heller, M. Klomfass, H. Neuberger and P.
Vranas, FSU--SCRI--92C--150/RU--92--43, to appear in the proceedings
of the {\sl LATTICE '92} conference in Amsterdam.

\refis{SWENDSEN} A.M. Ferrenberg and R.H. Swendsen, \prl 61, 1988,
2635, [erratum 63, (1989), 1658]; \prl 63, 1989, 1196.

\refis{HELLN} U. M. Heller, H. Neuberger, \pl 207B, 1988, 189; H.
Neuberger, \prl 60, 1988, 889.

\refis{MK} M. Klomfass, to appear in the proceedings of the {\sl
LATTICE '92} conference in Amsterdam, and in preparation. See also
Ph.D. thesis, FSU--SCRI--92T-117 (1992).

\refis{ZIMMERW} F. Zimmermann, J. Westphalen, M. G\"{o}ckeler, H. A.
Kastrup, talk presented by F. Z. at the International Symposium
Lattice '92, Amsterdam, Sept. 15--19, 1992; to appear in the
proceedings.

\refis{LUSCHER} M. L\"{u}scher, \np B354, 1991, 531; \np B364, 1991,
237.

\refis{LWFF} M. L\"{u}scher, U. Wolff, \np B339, 1990, 222.

\endreferences

\if \epsfpreprint N \endpage

\head{Figure Captions.}

\halign{#\hfill\qquad &\vtop{\parindent=0pt \hsize=5in \strut#
\strut}\cr
Figure 2.1: & \captiontwoone \cr
&\cr
Figure 2.2: & \captiontwotwo \cr
&\cr
Figure 2.3: & \captiontwothree \cr
&\cr
Figure 3.1: & \captionthreeone \cr
&\cr
Figure 3.2: & \captionthreetwo \cr
&\cr
Figure 3.3: & \captionthreethree \cr
&\cr
Figure 3.4: & \captionthreefour \cr
&\cr
Figure 3.5: & \captionthreefive \cr
&\cr
Figure 3.6: & \captionthreesix \cr
&\cr
Figure A.1: & \captionAone \cr
}
\fi

\endit